\documentclass{aa}  

\usepackage{graphicx}
\usepackage{txfonts}
\usepackage{placeins}

\usepackage{multirow}
\usepackage{threeparttable}
\usepackage{hyperref}

\begin{document}

   \title{Chemical hints of Population III stars from silicon abundances in massive galaxies}

   \subtitle{}

   \author{E. Eftekhari\inst{1,2,3}\corrauth{eftekhari.eli@gmail.com}        
        \and A. Ferré-Mateu\inst{1,2,4}\email{aferre@iac.es}
        \and A. Vazdekis\inst{1,2}\email{alexandre.vazdekis@iac.es}
        \and F. La Barbera\inst{5}\email{francesco.labarbera@inaf.it}
        \and M. A. Beasley\inst{1,2,4}\email{beasley@iac.es}
        \and J. P. V. Benedetti\inst{1,2}\email{joao.pedro.benedetti@iac.es}
        \and M. Kriek\inst{3}\email{kriek@strw.leidenuniv.nl}
        }

   \institute{Instituto de Astrof\'isica de Canarias, Calle Vía Láctea S/N, E-38205, La Laguna, Tenerife, Spain
   \and Departamento de Astrofísica, Universidad de La Laguna, 38206, La Laguna, Tenerife, Spain
   \and Leiden Observatory, NL-2300 RA, P.O. Box 9513, Leiden, The Netherlands
   \and Centre for Astrophysics and Supercomputing, Swinburne University, John Street, Hawthorn VIC 3122, Australia
   \and INAF-Osservatorio Astronomico di Capodimonte, sal. Moiariello 16, I-80131, Napoli, Italy }

   \date{}

\abstract{The formation of massive galaxies remains a fundamental question in astrophysics, with recent JWST observations suggesting that intense star formation occurred earlier than previously expected. We used the prototypical massive relic galaxy NGC\,1277 as a fossil record to probe the chemical signatures of early star formation.  With a stellar mass of $1.2\pm0.4\times10^{11}$M$_{\odot}$ and a compact structure (half-light radius of $1.2$\,kpc), NGC\,1277 is representative of massive relic galaxies—systems that formed rapidly at high redshifts (`red nuggets' at $z > 2$) and have since undergone largely passive evolution, preserving the imprint of their earliest stellar populations. Using deep near-infrared spectroscopy, we identified unusually strong silicon (Si) absorption features in this galaxy.  We measure [Si/Fe]=$1.05^{+0.45}_{-0.27}$ and [Mg/Fe]=$0.36^{+0.27}_{-0.24}$, corresponding to a significantly enhanced silicon-to-magnesium ratio ([Si/Mg]=$0.67^{+0.45}_{-0.27}$), which cannot be reproduced by current stellar population models or typical core-collapse supernova yields. This enhancement suggests a contribution from very massive, metal-poor progenitors during the earliest phases of star formation. Such conditions are consistent with nucleosynthetic yields from pair-instability supernovae or other enrichment channels associated with extremely massive stars, whose chemical signatures are expected to remain less diluted in systems like NGC\,1277 due to their rapid formation timescales and minimal late-time accretion. This result provides rare chemical evidence of early massive-star enrichment preserved in a local relic galaxy and opens a new observational window into the chemical evolution of the Universe.}


   \maketitle
   
\nolinenumbers

\section{Introduction}\label{sec1}

Recent discoveries with the \textit{James Webb} Space Telescope (JWST) have revolutionised our understanding of the early formation of massive galaxies, revealing massive, quiescent red galaxies at unexpectedly high redshifts ($z\sim$7-10; \citealt{duan2024, weibel2025}). These findings suggest that the first phase of massive galaxy formation may have occurred much earlier than previously thought, potentially pushing the initial burst of star formation into an epoch when the first generation of stars may have contributed significantly to the chemical enrichment of the Universe. This emerging picture aligns with the increasingly accepted two-phase formation scenario for massive galaxies~\citep{oser2010, vandokkum2010}. In this modern framework, the first phase involves a rapid and intense formation of compact, massive galaxies, commonly referred to as `red nuggets', occurring by $z > 2$~\citep{damjanov2009}. Following this brief ($<$1~Gyr) initial phase, these systems undergo substantial evolutionary growth over cosmic time, primarily in size, and to a lesser extent in mass, through mergers and gas accretion. The giant early-type galaxies (ETGs) observed in the local Universe are the massive end products of this process. 
Crucially, the compact red nugget formed during the initial phase with the earliest generation of stars is not destroyed but can survive, remaining buried in the central regions of ETGs. Subsequent star formation from recycled and accreted material was layered over this core, making it increasingly difficult to disentangle and reconstruct the pristine chemical conditions of the first phase of galaxy formation.

Because merger-driven evolution in the second phase is a stochastic process, it is believed that, in very rare instances, some of these high-\textit{\textit{z}} red nuggets remained untouched, mostly avoiding the subsequent accretion that characterises this second phase. They thus survived in their compact state until nowadays~\citep{naab2009}. These ancient, undisturbed systems are known as `massive relic galaxies' or `local red nuggets'~\citep{trujillo2009, ferre2017, spiniello2024}. What makes these galaxies particularly valuable is that they retain the structural and dynamical properties of their high-redshift counterparts, preserving key imprints of their early formation history. 

Beyond their structural properties, massive relic galaxies offer a rare window into the chemical enrichment processes that took place during the earliest phases of galaxy buildup. Because their stellar populations formed rapidly and have remained largely undisturbed, their chemical abundances can retain signatures of the nucleosynthetic processes that took place in the earliest stage of their formation. In this sense, relic galaxies can be viewed as `chemical fossils', allowing us to probe the nature of the first generations of stars through their imprint on later stellar populations.

Identifying relic systems is challenging due to their extreme rarity~\citep{quilis2013}, but a handful of confirmed examples exist. Among them, NGC\,1277 stands out as the prototypical case. First noted for hosting one of the most massive supermassive black holes known~\citep{vandenbosch2012, ferre2015},  NGC\,1277 is a massive (M$_{\ast}=1.2\pm0.4\times10^{11}$M$_{\odot}$), extremely compact (R$_{\rm e}=1.2$\,kpc), rapidly rotating system whose internal structure and stellar population properties closely resemble the compact objects at \textit{z} $\sim2$~\citep{trujillo2014, ferre2017}. Surprisingly, it is strikingly deficient in dark matter~\citep{comeron2023}. The unique evolutionary history of NGC\,1277 is further supported by its remarkably old and metal-rich globular cluster system, which lacks the typical blue, metal-poor subpopulation often associated with later accretion events~\citep{beasley2018}. 

Chemical abundance ratios provide key insights into the star formation histories (SFHs) of galaxies. Since different elements are produced by stars of varying masses and lifetimes, their abundance ratios can act as cosmic clocks, tracing the timescales of star formation~\citep{carretero2004, baumgartner2005}. In particular, $\alpha$-elements (such as Mg and Si) are synthesised in massive stars and ejected through core-collapse supernovae (CCSNe), while Fe is primarily contributed by Type\,Ia supernovae (SNe\,Ia) over timescales $\gtrsim$~1 Gyr. Therefore, these abundance ratio measurements serve as a crucial diagnostic for understanding how rapidly a galaxy formed the bulk of its stellar populations. However, to date, the assessment of $\alpha$-enhancement in massive galaxies, in particular massive relic ones, has largely relied on measurements of Mg, which is the primary tracer used in optical studies. While Mg plays a key role, it is not the sole contributor to a galaxy's overall $\alpha$-enhancement. Moreover, it is well established that, unlike in the Milky Way, the various $\alpha$-elements in massive galaxies are not enhanced in lockstep \citep{vazdekis1997}. Given that different $\alpha$-elements are produced in different proportions by various nucleosynthetic channels and in stars of different masses, constraining multiple $\alpha$-elements is necessary for a deeper understanding of chemical enrichment history and the nature of the stellar populations responsible for it.

Unfortunately, constraining  certain $\alpha$-elements, such as Si or O, remains challenging in the optical spectral window, as there are no individual absorption features with strong sensitivity to these elements that can be properly measured (see e.g.~\citealt{johansson2012}). This limitation is particularly pronounced in pressure-supported systems such as massive galaxies, which exhibit high velocity dispersions that significantly broaden and smear the observed spectral features. A prime example is NGC\,1277, whose central region reaches a velocity dispersion of $\sigma \sim 440\,\mathrm{km\,s^{-1}}$~\citep{trujillo2014} . 

The near-infrared (NIR) spectral range offers a promising alternative, as it contains prominent absorption features for several $\alpha$-elements, including Si, as well as iron-peak elements. By exploiting these features, it becomes possible to access strong and relatively clean diagnostics of elements that are otherwise difficult to constrain in the optical, providing complementary insight into the chemical enrichment history of massive galaxies. In particular, silicon exhibits prominent absorption features in the NIR, making it especially well suited for abundance studies in this wavelength range. In this work, we used deep NIR spectroscopy to investigate the chemical abundances of the massive relic galaxy NGC\,1277. Our goal was to explore whether its abundance patterns provide insight into the nature of the earliest stellar populations and the chemical conditions present during the initial phase of galaxy formation.

This paper is organised as follows. In Sect.~\ref{sec2} we describe the observational data, including the NIR spectroscopy of NGC\,1277 and the comparison sample of massive ETGs. Section~\ref{sec3} presents the stellar population synthesis (SPS) models used to interpret the spectral features, while Sect.~\ref{sec4} introduces the supernova (SN) yield models adopted to explore the origin of the observed abundance patterns. The measurement of spectral indices is detailed in Sect.~\ref{sec5}, followed by the estimation of elemental abundances in Sect.~\ref{sec6}. Our main observational results are presented in Sect.~\ref{sec7}. In Sect.~\ref{sec8} we discuss the implications of the derived abundance ratios in the context of nucleosynthetic enrichment and galaxy formation, and we summarise our main conclusions in Sect.~\ref{sec9}.

\section{Samples and data}\label{sec2}

    \subsection{NGC\,1277}\label{sec21}

    For this study, we utilised deep spectroscopic observations of the prototypical massive relic galaxy NGC\,1277, targeting the J and H bands in the NIR. The data were obtained using the EMIR, the NIR multi-object spectrograph at the 10.4\,m Gran Telescopio de Canarias (GTC), under programme GTC3-17B (PI: M. Beasley). Observations were conducted over five nights in 2017, using a 0.6$^{\prime\prime}$ wide long-slit aligned along the major axis of the galaxy.

    Light was dispersed using the J and H grisms, covering wavelength ranges of 1.2-1.3\,$\mu$m and 1.5-1.8\,$\mu$m, with spectral resolutions of $R \sim 5200$ and $\sim4300$, respectively. An ABBA nodding pattern with A and B positions offset by 90$^{\prime\prime}$ was employed to enable accurate sky subtraction. Each 120\,s exposure was repeated across multiple cycles, resulting in 48 exposures in J (12 ABBA cycles) and 36 exposures in H (9 cycles). To correct for telluric absorption, A0V and F- and G-type standard stars were observed throughout the observing run. Telluric correction was then performed using the \textsc{Molecfit} software~\citep{smette2015,kausch2015}, which fits synthetic atmospheric transmission models to the observed stellar spectra. Telluric corrections were applied row by row, accounting for molecular absorption features such as H$_2$O, CO$_2$, and CH$_4$ (see Appendix~\ref{app:telluric} for more details).
    
    Due to degradation introduced by EMIR dome flats, flat-fielding was omitted. This had no impact on the analysis since the extraction focused on a small 18-pixel region around the photometric centre of the galaxy, where large-scale flat-field effects are negligible. Bad pixels and cosmic rays were removed using the \textsc{L.A.Cosmic} routine~\citep{vandokkum2001}. Wavelength calibration was performed using OH airglow lines, and in the J band, Xe arc lines were used to supplement calibration at wavelengths $<$12,000\,\AA\ where OH coverage was sparse. The spectra were geometrically rectified by aligning the spatial centroid of each column with the galaxy centre. Co-addition of the 2D spectra and standard deviation maps was performed using the IRAF \texttt{imcombine} task. The accuracy of the standard deviation maps was validated by measuring noise levels in continuum windows, yielding scale factors of 1.03 (J) and 1.00 (H), indicating consistency without the need for rescaling.
    
    Final 1D spectra were extracted symmetrically about the galaxy centre using a 18-pixel aperture (corresponding to 1\,$R_\mathrm{e}$, or 1.2\,kpc). We achieved signal-to-noise ratios (S/Ns) of $\sim$160\,\AA$^{-1}$ in the J band and $\sim$70\,\AA$^{-1}$ in the H band, which are sufficient for reliable stellar population analysis~\citep{eftekhari2021}. We derived the radial velocity and velocity dispersion of the extracted spectra within 1\,$R_\mathrm{e}$ using the \texttt{pPXF} code~\citep{cappellari2023} in combination with the E-MILES SPS models. These kinematic parameters were used to shift the spectra to the rest frame and to convolve the two bands to a common velocity dispersion of $\sigma = 440$\,km\,s$^{-1}$ (the highest value in the central regions). 
        
    \subsection{Massive ETGs}\label{sec22}
    As a comparison sample, we used observations of a sample of massive ETGs from \citet{labarbera2019}, consisting of seven nearby ETGs. All galaxies in this sample lie at the high-mass end of the ETG population, with central stellar velocity dispersions in the range of 280--320~km\,s$^{-1}$. Five are brightest cluster galaxies while two others are satellites. The data of these galaxies consist of high-S/N spectra ($>$ 170~\AA$^{-1}$ in optical)  from the X-shooter spectrograph on the Very Large Telescope (VLT) \citep{vernet2011}, covering a wide wavelength range from the ultraviolet (3000~\AA) to the NIR (2.5~$\mu$m), with a spectral resolution of $R \sim 5500$ in the NIR arm. 1D spectra were extracted along the slit after carefully rectifying and combining 2D exposures.

    To enable a direct comparison with NGC\,1277, we constructed a stacked spectrum from the sample with two different extractions: 
    one was extracted within an aperture of 1.2~kpc, i.e. matching the central aperture of NGC\,1277; the other was within 1~R$_{e}$. The effective radius adopted corresponds to the bulge component of the galaxies, with an average value of R$_{e} \sim$ 3.9~kpc (see Appendix~C of \citealt{labarbera2019}). To ensure a consistent spatial sampling, spectra were extracted within a width of 1.3" ($\sim$1.65~kpc) around the photometric centre of each galaxy, equivalent to 1.5 times the mean seeing full-width at half maximum of the observations. All extracted spectra were re-binned adaptively in the dispersion direction to achieve a uniform minimum S/N of 50~$\AA^{-1}$ across the NIR wavelength range, ensuring reliable measurement of absorption features.
    The resulting stacked spectrum serves as a representative benchmark for typical massive ETGs in the local Universe.

\section{Stellar population synthesis models}\label{sec3}

    \subsection{Updated E–MILES}\label{sec31}

    We employed the updated E-MILES SPS models (\citealt{rock2015, rock2016, vazdekis2016}; Vazdekis et al. in prep.) to interpret the observed spectral features. These models extend into the NIR by both the original \citep{cushing2004, rayner2009} and the Extended IRTF (E-IRTF) stellar library \citep{villaume2017}, covering wavelengths from 1680 to 50,000,\AA. These models utilise theoretical isochrones from the BaSTI \citep{pietrinferni2004} and Padova \citep{girardi2000} databases, which are converted into the observational plane using extensive photometric stellar libraries (see \citealt{vazdekis2016}). The NIR predictions are reliable for stellar populations older than 1\,Gyr. The spectral predictions of the SPS models have a resolution of R$\sim$1400-3500 in the optical and R$=$2000 in the NIR. The E-MILES models offer a range of initial mass functions (IMFs), including universal and revised Kroupa, Chabrier, and both unimodal (i.e. single power-law) and bimodal (i.e. low-mass tapered) IMFs. 
    
    In this study, we utilised the updated E-MILES model predictions based on the BaSTI isochrones, considering ages from 1 to 14 Gyr and total metallicities from solar up to the $+$0.26~dex. We adopted a low-mass tapered bimodal IMF with a logarithmic slope of 1.3, closely resembling a standard Kroupa IMF \citep{vazdekis1996}, as well as a bottom-heavy IMF with a slope of 3.0 to better match the IMF measured in the centres of massive galaxies \citep{labarbera2019} and of NGC\,1277 \citep{martin2015d}.
    
    \subsection{Conroy et al. (2018) models}\label{sec32}

    To quantify the sensitivity of spectral features to elemental abundances, we used the stellar population models of \citet{conroy2018}. These models provide flexible and comprehensive predictions of stellar spectra across a range of ages between 1 and 13.5~Gyr and metallicities between [Fe/H]$=-$1.5 and $+$0.3\,dex.
    
    The \citet{conroy2018} models use empirical spectral libraries (based on MILES and IRTF) and theoretical isochrones (MIST; \citealt{choi2016, dotter2016}) to generate synthetic spectra and adopt a Kroupa IMF. These models cover a broad wavelength range, extending from 0.37 to 2.4~$\mu$m, and include detailed treatment of various stellar atmospheric parameters and elemental abundances. In particular, they allow for variations in individual element abundances, including $\alpha$-elements and other important contributors to stellar opacity and absorption features. The publicly available model simple stellar population (SSP) spectra are provided at a spectral broadening of $\sigma \sim 100$\,km\,s$^{-1}$
    
    In this study, we used the models with an age of 13~Gyr and metallicity [Fe/H] $=+$0.2\,dex, exploring three different abundance patterns: (i) models with [Mg/Fe] $=\pm0.3$\,dex while keeping other elements fixed at solar values, (ii) models with [Si/Fe] $=\pm0.3$\,dex with all other elements fixed, and (iii) solar-scaled models.

\section{Supernova yield models}\label{sec4}

To interpret the observed [Si/Mg] abundance ratios in NGC\,1277 and in the sample of local massive ETGs, we compared our measurements with theoretical SN nucleosynthetic yields as a function of progenitor mass and metallicity. These yield models are taken from the compilation of \citet{nomoto2013}, and include predictions for several SN types, 
each contributing differently to the chemical enrichment of galaxies. Their relative yields of Si and Mg provide key diagnostics for the origin of the abundance patterns observed.

\begin{itemize}
    \item CCSNe: These explosions originate from stars in the mass range 10--40\,M$_\odot$, and their nucleosynthesis patterns depend strongly on the progenitor mass and initial metallicity. These yields are taken from \citet{kobayashi2006}.

    \item Hypernovae (HNe): These energetic explosions of massive stars ($>$25\,M$_\odot$) with significantly larger explosion energies than standard CCSNe produce enhanced yields of $\alpha$-elements such as Si and Mg. The HN yields used in our work are based on the models of \cite{umeda2002b, umeda2005}, and are included in the \cite{kobayashi2006} framework.

    \item Pair-instability supernovae (PISNe): These explosions result from very massive metal-free stars (140-260\,M$_\odot$) and are characterised by an overproduction of intermediate-mass elements, particularly Si. The PISN yields adopted in this work are from \cite{umeda2002a}. These models are particularly relevant for probing the chemical enrichment by Population III (Pop~III) stars.

\end{itemize}

\begin{table*} 
        \centering
        \caption{Definition of spectral indices.}
                
        \label{tab:tab1} 

        \begin{tabular}{lccccccccr} 
                \\
                \hline
            Index & $\lambda_{blue1}$ & $\lambda_{blue2}$ & $\lambda_{center1}$ & $\lambda_{center2}$ & $\lambda_{red1}$ & $\lambda_{red2}$ & Reference \\
            & (\AA)   & (\AA) & (\AA) & (\AA) & (\AA) & (\AA) &  \\
            \hline
            MgI1.18 & 11740 & 11810 & 11810 & 11850 & 11905 & 11935 & \cite{eftekhari2021}$^\ast$ \\
            SiI1.20 & 11900 &  11940 &  11955 &  12005 &  12135 &  12165 & \cite{rock2015phd}$^\ast$ \\
            SiI1.59 & 15840 & 15860 & 15870 & 15910 & 15925 & 15950 & \cite{gasparri2021}$^\ast$ \\
                \hline
        \end{tabular}

    \begin{tablenotes}
    \footnotesize
    \item[*] * The wavelengths are adjusted.
    \end{tablenotes}

\end{table*}

The SN yield metallicities used in this study were: Z = 0.0000, 0.0010, 0.0040, 0.0080, 0.0200, and 0.0500. To classify SN types, we followed the progenitor mass and explosion energy criteria of \cite{nomoto2013}. Specifically, SNe were categorized as: (i) PISNe for progenitor masses $\geq$140~M$_\odot$; (ii) HNe for selected mass-energy pairs [20,10], [25,10], [30,20], [40,30], [100,60], with each pair denoting [mass in M$_\odot$, energy in $10^{51}$~erg]; and (iii) CCSNe for explosion energies of 1.0$\times$10$^{51}$~erg and progenitor masses $<$140~M$_\odot$ not falling into the HN combinations.  

For each SN model, total elemental yields were computed by summing over all stable isotopes. The [X/Fe] ratios were derived for each progenitor by comparing the logarithmic yield ratios, log(X/Fe) against solar reference values from \citet{asplund2009}.
These yield predictions provide a physically motivated framework to understand the origin of the observed elemental abundance ratios and to explore the possible contribution of different stellar populations to the earliest chemical enrichment.

\section{Spectral index measurement}\label{sec5}

To quantify the absorption strengths of NIR features, we measured Si and Mg spectral indices. The initial index definitions were drawn from \citet{eftekhari2021}, \citet{rock2015phd}, and \citet{gasparri2021}. However, due to specific artefacts in the observed spectra such as residuals from imperfect telluric correction, we refined several of these definitions to mitigate the impact of spectral shape peculiarities and local flux calibration uncertainties. The adopted index definitions used in this study are presented in Table~\ref{tab:tab1}. We verified that all clean, artefact-free definitions lead to consistent and robust measurements.

In particular, we modified the red pseudo-continuum band of the MgI\,1.18\,$\mu$m feature to avoid a structured bump and dip in the NGC\,1277 spectrum that distorted the local continuum estimate. Similarly, the red pseudo-continuum of the SiI\,1.20\,$\mu$m feature was redefined, as the original band from \cite{gasparri2021} was located too far from the absorption line centre, increasing susceptibility to systematic errors from broad-band flux calibration. To improve robustness, the blue pseudo-continuum band of the same index was also broadened to enhance the S/N and better anchor the local continuum level. The same blue pseudo-continuum adjustment was applied to MgI\,1.58\,$\mu$m.

To quantify uncertainties on the measured indices, we provided both the flux and corresponding error spectra as input to the \texttt{indexf} code\footnote{\url{https://indexf.readthedocs.io/}} \citep{cardiel2010}. The error estimates are derived through analytical propagation of flux uncertainties, following standard first-order error propagation. Specifically, the uncertainty in each index is calculated by considering both the noise in the flux within the feature band-pass and the propagated uncertainty in the interpolated pseudo-continuum. The latter is based on the uncertainties in the blue and red continuum regions used to estimate the local continuum level across the feature. All index values reported in this study are measured on flux-calibrated, rest-frame spectra smoothed to a common velocity dispersion of 440~km/s, matching the central velocity dispersion of NGC\,1277 (the highest in our galaxy sample) to ensure consistent comparison across models and datasets.
The resulting index measurements form the basis for the elemental abundance analysis in Sect.~\ref{sec6}.
    
\section{Elemental abundance estimation}\label{sec6}
    
We estimated [Si/Fe] and [Mg/Fe] for both NGC\,1277 and the sample of massive ETGs by using a set of four measured spectral indices: two NIR Si features (SiI1.20 and SiI1.58) and two Mg features (MgI1.18 and Mgb5177\footnote{Definition from \citet{worthey1994}.}) from both optical and NIR. Given the limited accuracy of current SPS models in the NIR, particularly for elemental abundance response functions, our goal is not to derive precise elemental abundances, but to provide a plausible range for [Si/Fe], [Mg/Fe], and their ratio. The NIR response models remain underdeveloped compared to those in the optical, and the absolute calibration of element-sensitive features in this regime is still uncertain. 

We used E-MILES SPS models and incorporated the abundance response functions from \citet{conroy2018}, which are computed assuming a Kroupa IMF, to simulate $\pm$0.3\,dex variations in [Si/Fe] and [Mg/Fe]. These responses were applied to E-MILES model grids with a bimodal IMF slope of 3.0 (representative of the IMF in the central regions of massive galaxies) to track how the four indices respond to abundance changes. For consistency, we adopted models with fixed stellar population parameters: age = 13\,Gyr and [M/H] = $+$0.2\,dex, and evaluated three abundance configurations: [Si/Fe] = $\pm$0.3\,dex, [Mg/Fe] = $\pm$0.3\,dex, and solar-scaled.

We caution that the abundance response functions used here are computed for abundance variations of only $\pm$0.3\,dex, whereas the Si-sensitive indices measured in NGC\,1277 require extrapolation beyond this range when interpreted under a linear-response approximation. The resulting absolute [Si/Fe] values should therefore be regarded as model-dependent estimates rather than precise abundance measurements. At such high inferred Si levels, the response of the Si features can become non-linear due to line saturation, changes in the pseudo-continuum, and opacity effects in cool stellar atmospheres. For this reason, our most robust conclusion is the differential one: NGC\,1277 exhibits significantly stronger NIR Si-sensitive absorption than the comparison massive ETG sample.

To estimate abundances, we employed two complementary techniques. With
Method 1, we estimated [Si/Fe], [Mg/Fe], and [Si/Mg] by projecting the observed indices onto a model response plane defined by solar-scaled, Si-enhanced, and Mg-enhanced predictions. A Monte Carlo sampling of $n=1000$ realisations was performed to determine the associated uncertainties, obtaining a mean [Si/Fe], [Mg/Fe], and [Si/Mg]. 

With Method 2, we adopted a local linear relation calibrated over the [$-0.3$, $0.0$, $+0.3$]\,dex abundance range to estimate the abundance values that best reproduce the observed line strengths. For NGC\,1277, this procedure involves extrapolation beyond the calibrated response-function range, particularly for the Si-sensitive indices. Error propagation was carried out analytically from the interpolation slope and index uncertainty.

Combining two Si indices (SiI1.20 and SiI1.58) with two Mg indices (MgI1.18 and Mgb5177), we formed four distinct Si–Mg index pairs. For each pair, we applied two independent estimation methods, resulting in a total of eight abundance measurements per galaxy for [Si/Fe], [Mg/Fe], and [Si/Mg]. These values are listed in Table~\ref{tab:tab2}.

To obtain representative abundance ratios from these independent determinations, we employed a Bayesian hierarchical model that accounts for both measurement uncertainties and possible intrinsic scatter among the estimates. In this framework, each measured value (and its associated uncertainty) is treated as an independent realisation drawn from a parent distribution characterised by a true mean and dispersion. The resulting posterior distributions provide a statistically consistent summary of the abundance estimates obtained from the different index combinations, while retaining the systematic caveat associated with the response-function extrapolation discussed above. The median values and 68\% credible intervals of these posteriors are reported in Table~\ref{tab:tab2}.

\begin{table*}[t]
\centering
\caption{Individual abundance measurements for NGC\,1277 and the massive ETG sample.}
\label{tab:tab2}
\begin{tabular}{llccccc}
\hline
Galaxy & Method & Si index & Mg index & $[\mathrm{Si/Fe}]$ & $[\mathrm{Mg/Fe}]$ & $[\mathrm{Si/Mg}]$ \\
\hline
\multirow{8}{*}{NGC\,1277}
& 1 & SiI1.20 & Mgb5177  & 0.77 & 0.27 & 0.49 \\
& 2 & SiI1.20 & Mgb5177  & 0.71 & 0.33 & 0.38 \\
& 1 & SiI1.20 & MgI1.18 & 0.85 & 0.45 & 0.40 \\
& 2 & SiI1.20 & MgI1.18 & 0.71 & 0.48 & 0.23 \\
& 1 & SiI1.59 & Mgb5177  & 1.70 & 0.21 & 1.49 \\
& 2 & SiI1.59 & Mgb5177  & 1.93 & 0.33 & 1.60 \\
& 1 & SiI1.59 & MgI1.18 & 1.55 & 0.44 & 1.11 \\
& 2 & SiI1.59 & MgI1.18 & 1.93 & 0.48 & 1.45 \\
\hline
\multicolumn{4}{l}{Posterior median $\pm$ 68\% credible interval}
& $1.05^{+0.45}_{-0.27}$ & $0.36^{+0.27}_{-0.24}$ & $0.67^{+0.45}_{-0.27}$ \\
\hline
\hline
\multirow{8}{*}{XSGs}
& 1 & SiI1.20 & Mgb5177  & 0.22 & 0.24 & $-0.02$ \\
& 2 & SiI1.20 & Mgb5177  & 0.10 & 0.26 & $-0.16$ \\
& 1 & SiI1.20 & MgI1.18 & 0.25 & 0.30 & $-0.05$ \\
& 2 & SiI1.20 & MgI1.18 & 0.10 & 0.31 & $-0.21$ \\
& 1 & SiI1.59 & Mgb5177  & 0.48 & 0.22 & 0.26 \\
& 2 & SiI1.59 & Mgb5177  & 0.60 & 0.26 & 0.34 \\
& 1 & SiI1.59 & MgI1.18 & 0.47 & 0.30 & 0.17 \\
& 2 & SiI1.59 & MgI1.18 & 0.60 & 0.31 & 0.29 \\
\hline
\multicolumn{4}{l}{Posterior median $\pm$ 68\% credible interval}
& $0.34^{+0.27}_{-0.25}$ & $0.27^{+0.26}_{-0.25}$ & $0.07^{+0.27}_{-0.25}$ \\
\hline
\end{tabular}

\tablefoot{Individual abundance measurements are derived from eight combinations of Si and Mg spectral indices and abundance-estimation methods. The posterior rows report the median values and 68\% credible intervals from the Bayesian hierarchical analysis, providing a statistical summary of the inferred abundance ratios for each galaxy.}
\end{table*}

\section{Results}\label{sec7}

\begin{figure*}[ht!] 
        \centering
        
        \includegraphics[width=\hsize]{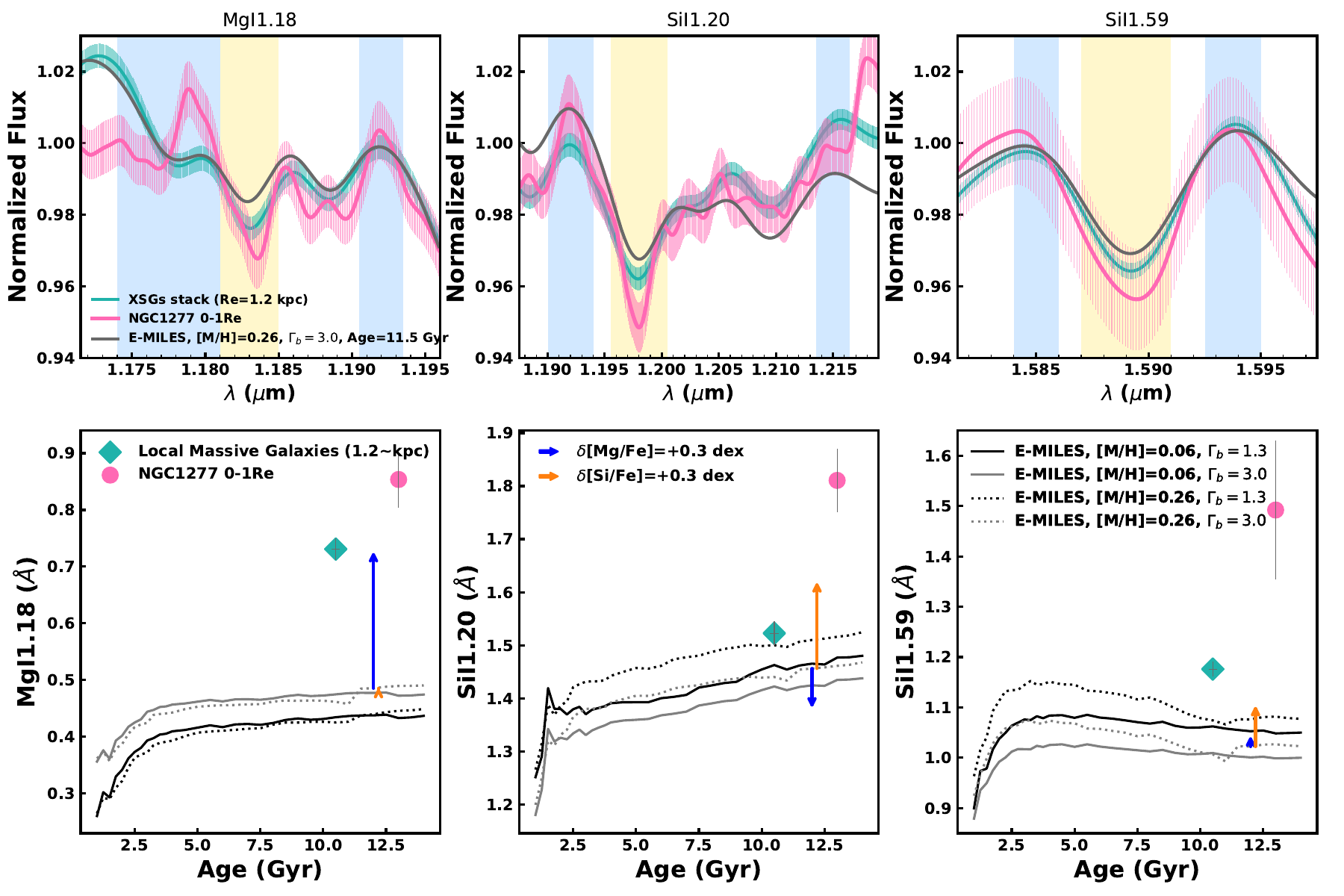} 
    \caption{NIR Mg and Si absorption line strengths in NGC\,1277 and massive ETGs. \textit{Top:} Zoomed-in view of the spectra of NGC\,1277 (pink line) and the stacked massive ETGs (cyan) around the MgI1.18, SiI1.20, and SiI1.59 features, with their corresponding error spectra. Yellow bands highlight these key absorption features, and light blue regions show the pseudo-continuum bands used for their index measurements. As reference, we show in grey an E-MILES model spectrum for an old, metal-rich, dwarf-enhanced population. \textit{Bottom:} Measured line-strength indices of MgI1.18, SiI1.20, and SiI1.59 plotted as a function of stellar population ages. Model predictions are shown for E-MILES assuming different IMF shapes (Milky Way-like in black, bottom-heavy in grey) and metallicities (solar as solid lines, super-solar as dotted lines). Arrows show index responses to $+$0.3\,dex enhancements in [Mg/Fe] (blue) and [Si/Fe] (orange) using response functions from \cite{conroy2018}, attached to models representing the average stellar population of massive galaxy centres. The observed values are overlaid as pink circles (NGC\,1277) and cyan diamonds (local massive ETGs). The Mg enhancement in NGC\,1277 confirms previous results in the optical, while the stronger Si absorption ($\sim$25\% higher than in typical ETGs) cannot be matched by current models, even when varying IMF or metallicity, suggesting a distinct chemical enrichment. }
        \label{fig:fig1}
\end{figure*}

We analysed key NIR absorption features of the massive relic galaxy NGC\,1277 and compared them with those of massive ETGs. The focus is placed on MgI1.18, SiI1.20, and SiI1.59 spectral indices, which are highlighted by yellow regions in the top row of Fig.~\ref{fig:fig1}, zooming into the spectra around the absorption lines. The spectrum of NGC\,1277 is shown  with its corresponding error spectrum to illustrate the observational uncertainties. We compare these absorption features with those from the stacked spectrum of a sample of local massive ETGs (within 1.2\,kpc to match the size of NGC\,1277). For comparison, predictions from updated E-MILES SPS models with old age, super-solar metallicity, and a dwarf-rich stellar IMF (following the results of \citealt{martin2015d}) are also shown in the top panels. The panels clearly show that the Mg and Si absorptions features in NGC\,1277 are deeper than both the local massive galaxies and the model. 

We measured the line-strength indices in all galaxies, as shown in the bottom panels of Fig.~\ref{fig:fig1}. The indices are presented as a function of age. We first confirm that the Mg line strength in NGC\,1277 is significantly stronger than in normal massive ETGs, consistent with optical results (lower-left panel;~\citealt{trujillo2014, martin2018}). However, a striking difference emerges when examining the Si indices (middle and right lower panels): this rare, ultra-compact massive relic galaxy exhibits exceptionally strong Si absorption, exceeding that of normal massive galaxies by approximately 25\%.

To explore the origin of this contrasting behaviour between the two $\alpha$-elements, we compared our measurements with SPS models: E-MILES models with a standard Kroupa-like IMF (i.e. a bimodal IMF with slope $\Gamma_b = 1.3$) and a dwarf-enriched (or bottom-heavy) IMF, implemented as a bimodal IMF with slope $\Gamma_b = 3.0$, where a larger fraction of stellar mass is in low-mass stars. We show the response of line-strength indices to different abundance variations based on the~\cite{conroy2018} models: (i) a $+$0.3\,dex enhancement in [Mg/Fe] and (ii) a $+$0.3\,dex enhancement in [Si/Fe]. Each arrow is placed at the model corresponding to the average stellar population parameters derived for their galaxy centres, illustrating the index variation. While the enhanced Mg line strength in NGC\,1277 can be attributed to its higher [Mg/Fe] ratio, which is robustly established through optical studies), the Si line strengths in NGC\,1277 cannot be explained by any modelled abundance variation\footnote{We also examined the response of these indices to all other elemental abundance variations available in the \citet{conroy2018} models. These responses are generally subtle compared to the observed offsets and to the Mg and Si response arrows, suggesting that contamination by another single element is unlikely to drive the strong Si absorption in NGC\,1277.}. Remarkably, this marks the first detection of such strong NIR Si absorption lines in a galaxy.

To assess the robustness of our results, we explored a range of potential systematic effects and alternative explanations. These include variations in the stellar population modelling, such as different IMF parameterisations beyond those considered in the main analysis, as well as the use of independent SPS models (E-MILES and X-shooter Spectral Library). We also inspected the behaviour of the relevant line strengths within the empirical stellar library feeding the models, tested the possible contribution of metal-poor populations and aperture effects. In addition, we compared our NIR-based abundance estimates with optical determinations of [Si/Fe] and [Mg/Fe] derived using independent methods. Overall, these analyses  support the robustness of the unusually strong NIR Si absorption in NGC\,1277 and its enhanced [Si/Mg] relative to massive ETGs. At the same time, the optical analysis yields a much less extreme [Si/Fe] enhancement, showing that the absolute Si abundance depends on the adopted spectral diagnostics. Our main conclusion is therefore that NGC\,1277 exhibits a distinctive Si-enhanced chemical signature most clearly revealed in the NIR, while the absolute [Si/Fe] value should be interpreted with caution (see Appendices~\ref{app:secA1}--\ref{app:secA4}  for details of these tests).

\section{Discussion}\label{sec8}

The elemental abundance pattern observed in NGC\,1277 reveals a clear departure from that of typical massive ETGs. In particular, the elevated [Si/Mg] ratio hints that silicon and magnesium, although both $\alpha$-elements, do not necessarily share the same enrichment history in this system, a difference that could be tied to the specific timing of star formation during its early stages.

The abundance estimates presented in Sect.~\ref{sec6} show that NGC\,1277 exhibits both enhanced [Si/Fe] and moderately elevated [Mg/Fe], with a clear excess of silicon relative to magnesium compared to massive ETGs. While these ratios provide complementary information, their difference ([Si/Mg]) offers a more direct diagnostic of the relative nucleosynthetic contributions from different enrichment channels. Because [Si/Mg] does not depend on the level of iron enrichment, it is largely insensitive to the delayed contribution of Fe from Type~Ia SNe, which operate on longer timescales. Instead, it isolates variations among $\alpha$-elements produced predominantly by massive stars, making it a more direct probe of the relative nucleosynthetic contributions from different high-mass progenitors. We therefore focused on [Si/Mg] as the key quantity for interpreting the origin of the observed abundance pattern.

The Si excess presents a challenge for current stellar population models, which fail to reproduce such a pattern. Since different types of SNe yield different ratios of $\alpha$-elements depending on their progenitor mass, explosion energy, and metallicity, we turned to nucleosynthetic yield predictions as a physically motivated framework to trace the likely origin of the observed abundance pattern. We thus estimated the relative abundance of [Si/Mg] in both NGC\,1277 and massive ETGs, and compared these values with the predicted yields from CCSN, HN, and PISN models. These events represent the dominant enrichment channels for massive stars at different initial mass regimes and metallicities.

\begin{figure}[ht!] 
        \centering
        \includegraphics[width=\hsize]{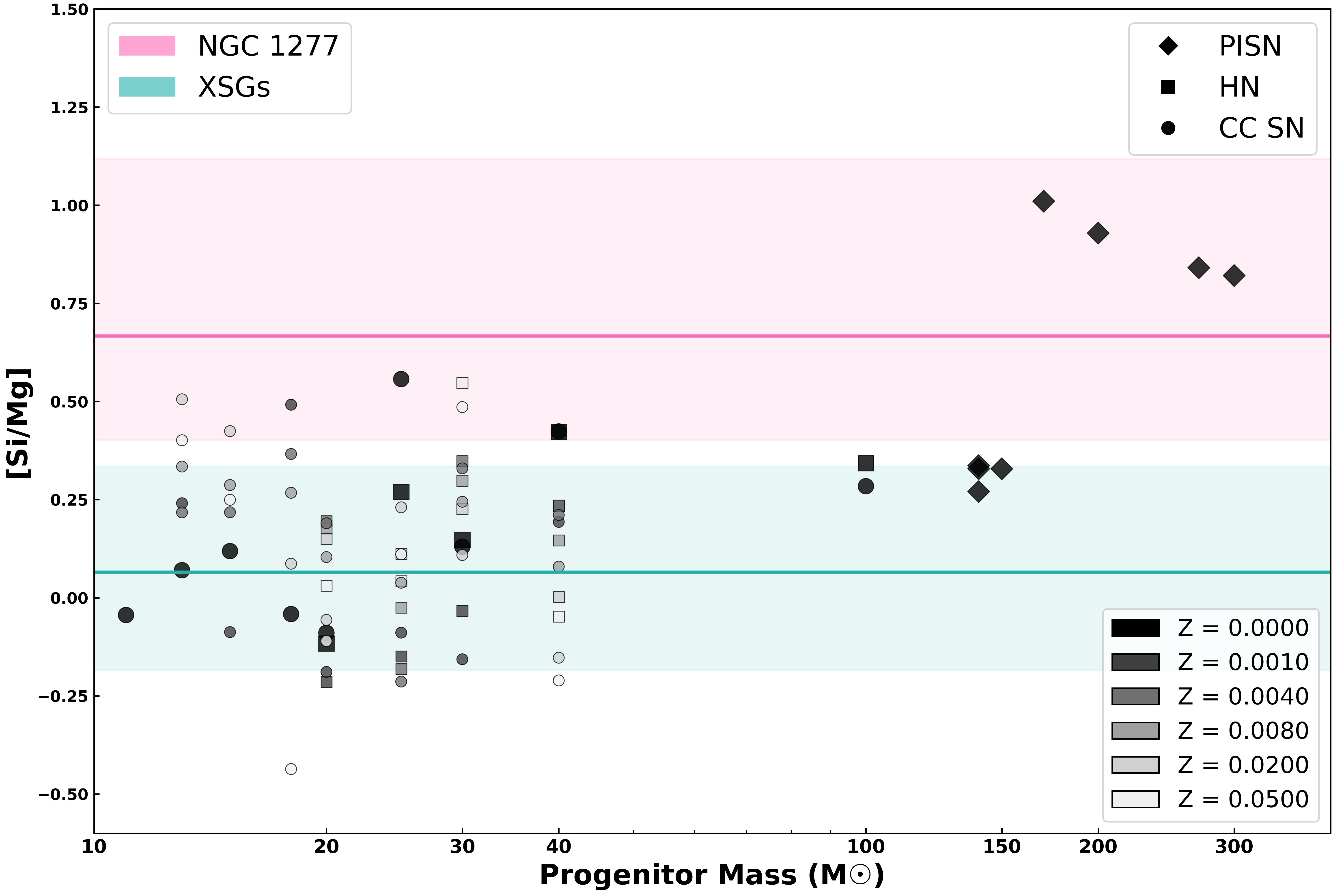} 
        \caption{[Si/Mg] as a function of SN progenitor mass.The measured [Si/Mg] abundance ratios are compared to theoretical nucleosynthetic yields from SN models across a range of progenitor masses.  Circles, squares, and diamonds denote the predicted yields for CCSNe, HNe, and PISNe, respectively, based on \cite{nomoto2013}.  The shaded bands represent the 68\% credible intervals of the inferred [Si/Mg] ratios for NGC\,1277 (pink) and the sample of local massive ETGs (cyan), derived from the Bayesian hierarchical analysis (Table~\ref{tab:tab2}). The solid lines mark the corresponding posterior medians.} 
        \label{fig:fig2} 
\end{figure}

Figure~\ref{fig:fig2} shows [Si/Mg] as a function of the main-sequence mass of SN progenitors, plotted on a logarithmic scale to better visualise the wide mass range of stellar masses spanning CCSNe, HNe, and PISNe; the progenitor metallicity ranges from metal-free (Z\,$=0$) to super-solar values (Z\,$=0.05$). According to \cite{heger2010}, stars more massive than $\sim$140M$_\odot$ are expected to explode as PISNe, while lower-mass stars undergo CC. These two SN types have markedly different nucleosynthetic yields, as illustrated in this figure.

We also show the 68\% credible intervals of the inferred [Si/Mg] ratios for NGC 1277 and the massive ETGs sample, respectively, derived from our Bayesian hierarchical analysis (Table~\ref{tab:tab2}). Despite the larger spread in the NGC\,1277 values, the two regions are clearly separated, indicating a distinct chemical signature for NGC\,1277. The [Si/Mg] values observed in NGC\,1277 partially overlap with yields from CCSNe (Z\,$=0.00-0.05$), HNe (Z\,$=0.000$, 0.050), and PISNe,  although all of these channels show significant scatter and their theoretical yields carry substantial uncertainties. 

Importantly, when considered as a function of progenitor mass and metallicity, only yields associated with the most massive, metal-poor progenitors consistently extend into the upper end of the observed [Si/Mg] distribution. Theoretical studies show that such extreme [Si/Mg] ratios arise under specific explosion conditions: PISNe from very massive stars are predicted to produce large amounts of intermediate-mass elements such as Si relative to Mg, often reaching [Si/Mg] $\gtrsim 0.5$ \citep{takahashi2018}, while more energetic explosions of massive stars (HNe) can also enhance Si through explosive nucleosynthesis, albeit typically over a more limited range \citep{heger2010}. In contrast, metal-rich CCSNe and HNe predominantly yield near-solar [Si/Mg] ratios, similar to those inferred for typical massive ETGs.

Taken together, these results suggest that, while multiple enrichment pathways may plausibly contribute to the chemical evolution of NGC\,1277, the high [Si/Mg] values require a non-negligible contribution from very massive progenitors characterised by disproportionately high silicon production relative to magnesium. In the nucleosynthetic yield models adopted here, such progenitors are associated with PISNe.

For completeness, we also examined the behaviour of [Si/Fe] and [Mg/Fe] separately as a function of progenitor mass (Appendix~\ref{app:sm_f}). While these quantities follow similar overall trends, they show larger degeneracies between different enrichment channels, further supporting that [Si/Mg] provides a cleaner diagnostic of the relative contribution of massive progenitors.

\subsection*{Implications for early chemical enrichment}
\begin{figure*}[ht!] 
        \centering
        \includegraphics[width=0.85\textwidth]{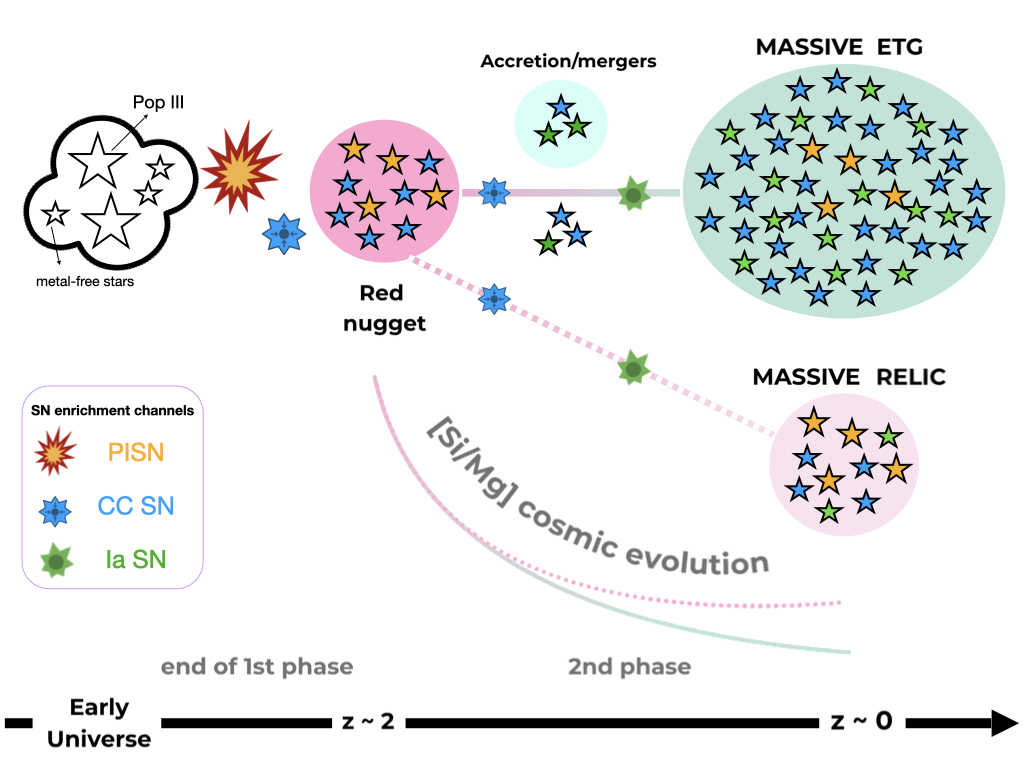} 
        \caption{Schematic illustration of early chemical enrichment and galaxy evolution under the two-phase formation scenario. In the early Universe (top left), metal-free stars, including both massive Pop~III stars (large stars) and less massive ones (smaller stars), form from pristine gas. Pop~III stars end as PISNe (orange bursts), while lower-mass metal-free stars produce CCSNe (blue bursts). The stars that form from this enriched material inherit the colour of the enriching SN type. They populate an early compact massive galaxy, or `red nugget' (pink background), with high [Si/Mg]. In normal massive galaxies, a second phase brings mergers and accretion, together with additional enrichment by CCSNe and SNe Ia (blue and green bursts), significantly diluting [Si/Mg]; they then evolve into massive ETGs (greenish background). In contrast, relic galaxies like NGC\,1277 do not experience significant mergers over the second phase, passively evolving with only minor additional enrichment (the lighter pink background exemplifies such a minimal dilution). The bottom-middle  tracks show the schematic [Si/Mg] evolution for each formation pathway.}
        \label{fig:fig3} 
\end{figure*}

The distinctive abundance pattern observed in NGC\,1277 is best understood within the two-phase galaxy formation scenario for massive galaxies. As illustrated schematically in Fig.~\ref{fig:fig3}, the early Universe is populated by metal-free stars. The most massive of these, Pop~III stars are expected, in current models, to explode as PISNe, while lower-mass metal-free stars produce CCSNe. These explosions enrich the ISM, forming new generations of stars. Massive galaxies thus start forming rapidly and very early on from very massive, metal-poor or even metal-free progenitors. During these initial stages, the ISM is enriched primarily by very massive, metal-poor progenitors—well represented by PISN models—yielding exceptionally high [Si/Mg]. As the first phase continues, CCSNe and HNe begin to contribute to the chemical enrichment, gradually diluting the [Si/Mg] ratio, even before external processes start occurring in the second phase. During this second phase, marked by accretion and merger events, the stars are being incorporated and enriched by metal-rich CCSNe and Type Ia SNe, further diluting [Si/Mg] (shown by the gradient line evolving from the red nugget phase to a regular massive ETG). 

However, NGC\,1277 follows a different enrichment track. Firstly, its  extremely intense initial star formation rates ($\sim$1000~M$_{\odot}$/yr; \citealt{trujillo2014}) made the early enrichment of [Si/Mg] particularly pronounced. This was followed by a rapid decline in the star formation rate, which minimised the dilution compared to the normal track. Thus, when NGC\,1277 reached its red nugget stage, it already presented more elevated [Si/Mg]. Given its relic nature, NGC\,1277 then evolves over a quiescent track (dotted line), experiencing only passive evolution, with minimal accretion over the second phase. The colour gradient from dark to lighter pink conveys this subtle chemical evolution, in contrast to the drastic [Si/Mg] flattening shown for massive ETGs.

Thanks to its unique relic nature, the stars we observe in NGC\,1277 today predominantly formed during the earliest, most intense phase of its SFH. As a result, their chemical abundance pattern retains the imprint of enrichment from very massive progenitors formed at the earliest stages, before subsequent chemical dilution. The exceptionally high [Si/Mg] values seen in NGC\,1277 indicate a galaxy that formed the majority of its stars extremely rapidly and experienced little subsequent chemical evolution in comparison to typical massive ETGs. Notably, such a short star formation timescale ($\lesssim$1\,Gyr) is comparable to those inferred for the compact quiescent galaxies now being identified at $z > 3$ with JWST \citep{carnall2024}. This fossil-like preservation of NGC\,1277's chemical signature is what makes it such a powerful probe of primordial stellar populations, as summarised in Fig.~\ref{fig:fig3}: a single pathway branching early, with relic galaxies like NGC\,1277 preserving the memory of their first-generation progenitors, while typical massive ETGs accumulate layers of chemical complexity over time.

We thus conclude that the distinctive chemical properties observed today in NGC\,1277 result from both the timing and intensity of star formation and the subsequent modest evolution. These considerations highlight that the distinctive [Si/Mg] signature in NGC\,1277 likely traces nucleosynthetic yields from very massive stars ($>$150M$_\odot$), which produce Si in disproportionately large amounts compared to Mg. This strongly supports a scenario in which the chemical composition of NGC\,1277 was imprinted by the earliest, most massive stars. While the exact identity of these progenitors remains uncertain, several early-Universe channels associated with extremely massive, metal-poor stars may contribute, including PISNe and related pair-instability outcomes, such as pulsational pair-instability events and near-threshold massive star endpoints \citep{heger2002, heger2010, woosley2007, whalen2014}. PISN models provide a well-developed theoretical framework that naturally reproduces the high [Si/Mg] ratios observed here. Regardless of the specific explosion mechanism, the observed abundance pattern in NGC\,1277 provides compelling evidence that the chemical composition of this galaxy was imprinted by the earliest and most massive stars. This represents the first identification of such a preserved [Si/Mg] signature in such massive galaxies.

While our analysis relies on stellar population models with a dwarf-rich IMF, which is appropriate for describing the observed stellar populations in NGC\,1277, the derived chemical pattern is best understood in the context of a time-varying IMF. In this scenario, an initial top-heavy phase dominated by massive stars, which include Pop~III and earliest generations of stars contributing to the enrichment of the interstellar medium (ISM), followed by a transition to a bottom-heavy mode~\citep{vazdekis1996,vazdekis1997,jerabkova2018,yan2019}. 

To further assess the plausibility of this picture, we computed IMF-weighted nucleosynthetic yields across a range of power-law IMF slopes. For a standard Salpeter IMF ($\alpha = 2.35$), the integrated [Si/Mg] ratio is $\sim$0.05\,dex, while in the extreme case of a top-heavy IMF ($\alpha = 0$), the ratio increases to $\sim$0.33,dex. Within the framework of the yield models explored here, very high [Si/Mg] values are attained only when the contribution of the most massive progenitors—represented by PISN yields—is combined with a top-heavy IMF during these first stages of galaxy evolution. Thus, the [Si/Mg] excess in NGC\,1277 reflects the chemical memory of its earliest, most massive stellar generations. A fully quantitative assessment of whether such an early very-massive-star contribution can reproduce both the enhanced [Si/Mg] ratio and the overall high metallicity of NGC\,1277 requires dedicated chemical-evolution modelling; we are currently developing such models, which will be presented in a follow-up study.

The unusually high [Si/Fe] and [Si/Mg] ratios found here should also be viewed in the broader context of the well-known abundance and IMF anomalies in the centres of massive ETGs. These systems have been shown to host bottom-heavy IMFs and strong Na-sensitive absorption features, often interpreted as very high [Na/Fe] (e.g. \citealt{labarbera2017, rock2017, alton2018}). Recent work has further highlighted the connection between Na enhancement and IMF variations in massive ETGs, reinforcing the view that the central regions of massive galaxies formed under extreme physical conditions, likely involving very rapid, high-density star formation and non-solar abundance patterns. However, we caution against interpreting Na as a direct tracer of the same enrichment channel responsible for the high [Si/Mg] ratio. Unlike Si and Mg, Na is an odd-Z element whose production is strongly sensitive to progenitor metallicity and neutron excess, and it may also receive contributions from asymptotic giant branch stars (e.g. \citealt{kobayashi2006}).

The unusually high [Si/Mg] in NGC\,1277 also aligns with predictions from theoretical and observational studies of early enrichment scenarios. \cite{takahashi2018} modelled non-rotating Pop~III progenitors and found that [Si/Mg] ratios in the range 0.33–0.92\,dex can result from zero-metallicity SNe, consistent with our derived values in NGC\,1277 ($0.67^{+0.45}_{-0.27}$; see Table~\ref{tab:tab2}). Additionally, \citet{sodini2024} analysed the chemical composition of neutral gas in damped Lyman-$\alpha$ systems at $z\sim$6, finding signatures of early silicon enrichment in the ISM of primordial galaxies. Their study provides independent gas-phase evidence of early Pop~III nucleosynthesis. 

Although our analysis focuses on Mg and Si, other elemental abundances (e.g. Al and Zn) are also known to carry additional information on chemical enrichment. In particular, abundance patterns involving odd- and even-Z elements can provide further constraints on nucleosynthetic pathways and the nature of the enriching sources. However, their measurement in the NIR is currently subject to larger modelling uncertainties and degeneracies. A comprehensive multi-element analysis, combining optical and NIR diagnostics, is beyond the scope of this work and will be addressed in future studies.

\section{Conclusions}\label{sec9}

We have presented a detailed analysis of NIR spectral features in the massive relic galaxy NGC\,1277 and compared them with those of a representative sample of massive ETGs. Our main findings can be summarised as follows:

\begin{itemize}

\item NGC\,1277 exhibits significantly stronger silicon absorption features than typical massive ETGs, with Si line strengths exceeding those of the comparison sample by $\sim25\%$.

\item The derived abundance ratios for NGC\,1277 indicate enhanced [Si/Fe] ($1.05^{+0.45}_{-0.27}$) and moderately elevated [Mg/Fe] ($0.36^{+0.27}_{-0.24}$), resulting in a high [Si/Mg] ratio ($0.67^{+0.45}_{-0.27}$) that is clearly distinct from that of normal massive ETGs.

\item The observed Mg and Si absorption features in NGC\,1277 are not readily explained by standard chemical enrichment scenarios, indicating a distinct abundance pattern compared to typical massive ETGs.

\item Comparison with SN yield models indicates that the high [Si/Mg] values are consistent with enrichment involving a significant contribution from very massive progenitors ($>$150~M$_\odot$), although multiple enrichment channels likely contribute.

\item The observed abundance pattern can be understood in the context of the relic nature of NGC\,1277, whose rapid and early SFH, combined with the absence of significant late-time accretion,  allowed it to retain chemical signatures that are significantly diluted in more typical massive galaxies.

\end{itemize}

Our discovery of an unusually strong Si absorption in NGC\,1277, reported in this work for the first time, provides a rare opportunity to probe the conditions of early galaxy formation. This is accessible not through direct observations at extreme redshifts, but through the fossil record preserved in local massive relics. While recent JWST observations have revealed massive, quiescent galaxies at unexpectedly high redshifts ($z\sim$7-10), the stellar populations in these early systems are too young for a reliable measurement of detailed chemical abundances, even with very high-quality data. In contrast, massive relic galaxies like NGC\,1277 contain evolved stellar populations that encode the chemical signatures of primordial star formation episodes. Measuring detailed abundance ratios in relics not only constrains early star formation conditions, but also provides observational leverage to estimate the initial masses of Pop~III stars, whose predicted yields, especially of Si, are highly mass-dependent~\citep{heger2002, beasley2003}. These results highlight the utility of NIR spectroscopy in unveiling the elemental imprints of early cosmic history. By using NIR spectral indices to trace elemental abundances in local massive relic galaxies, our work opens a new pathway to study the earliest stellar generations and chart the enrichment history of the cosmos.

\begin{acknowledgements}
      E.E., A.F.M., A.V. and J.P.V.B. acknowledge funding from the MCIN/AEI and the European Regional Development Fund (ERDF) through the POKEBOWL project PID2021-123313NA-I00. E.E., A.F.M. A.V., M.A.B. and J.P.V.B. acknowledge support from grants PID2022-140869NB-I00 and PID2024-162088NB-I00 from the Spanish Ministry of Science, Innovation and Universities (MCIU). This work has also been supported through the IAC project TRACES, partially funded by the state and regional budget of the Consejería de Economía, Industria, Comercio y Conocimiento of the Canary Islands Autonomous Community. F.L.B., A.V., and E.E. acknowledge support from the INAF mini-grant 1.05.23.04.01 and A.FM. has received support from RYC2021-031099-I of MICIN/AEI/10.13039/501100011033/ UE NextGenerationEU/PRTR. J.P.V.B. received the support of a fellowship from the ``la Caixa" Foundation (ID 100010434). The fellowship code is LCF/BQ/DI23/11990084. E.E. and M.K. acknowledge funding from the Dutch Research Council (NWO) through the award of the Vici grant VI.C.222.047.
\end{acknowledgements}

\bibliographystyle{bibtex/aa} 
\bibliography{NGC1277}

\begin{appendix}

\section{Telluric correction}\label{app:telluric}
To correct for telluric absorption, we used the software \textsc{Molecfit} \citep{smette2015, kausch2015}, which constructs a synthetic telluric transmission model directly from the science spectrum. Given that the spectral resolution of EMIR varies slightly with wavelength, we ran \textsc{Molecfit} on multiple spectral windows across the J and H bands and then combined the resulting models to produce a full telluric correction. This method allowed us to accurately remove absorption features from atmospheric H\textsubscript{2}O, O\textsubscript{2}, and CH\textsubscript{4}, which are particularly prominent in the NIR. The correction was applied row-by-row on the 2D galaxy spectra prior to 1D extraction. We verified the performance of the correction by comparing spectra before and after \textsc{Molecfit} application, confirming that telluric features were removed to better than the 2\% level across both bands (see \citealt{eftekhari2021phd}).

\section{Detailed look at the stellar library}\label{app:secA1}

\begin{figure*} 
        \centering
        \includegraphics[width=0.8\textwidth]{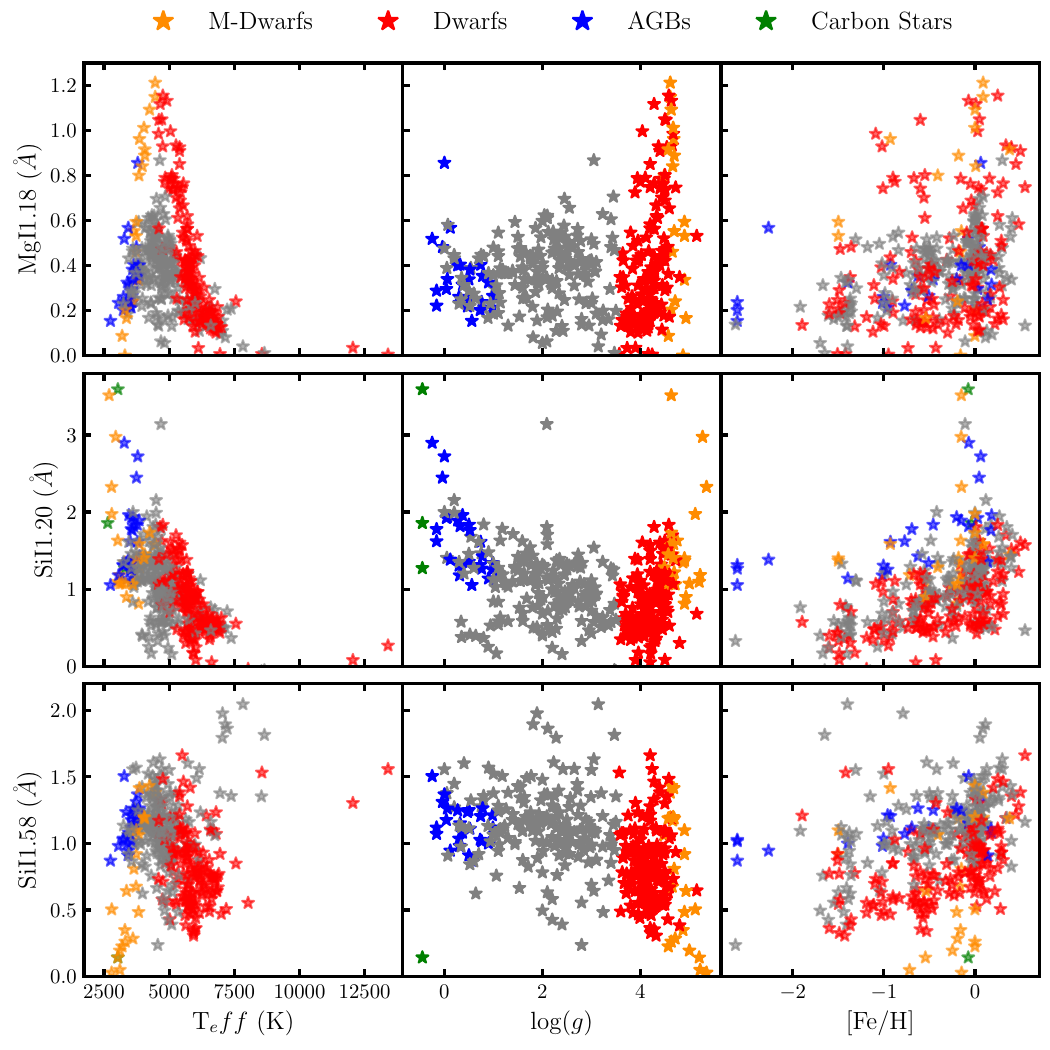} 
        \caption{Line-strength measurements for MgI1.18, SiI1.20, and SiI1.58 index dependences across the E-IRTF stellar library as functions of stellar effective temperature (T$_{eff}$), surface gravity (log($g$)), and metallicity ([Fe/H]). Different stellar types are colour-coded (see the legend).}
        \label{fig:fig6} 
\end{figure*}

To investigate the source of the discrepancies between SPS models and observations, we examined the stellar library used for E-MILES and \cite{conroy2018} SPS modelling, i.e. E-IRTF\footnote{Note that the \cite{conroy2018} model does not include the common stars among the IRTF and E-IRTF libraries while E-MILES uses all stars in the E-IRTF library.}. Figure~\ref{fig:fig6} displays the line-strength values of MgI1.18, SiI1.20, and SiI1.58 indices as functions of stellar effective temperature (T$_{eff}$), surface gravity (log($g$)) and metallicity ([Fe/H]). The stellar types (for stars cooler than 3900\,K) are drawn from Table\,2 of \cite{rock2015}. M-dwarf stars are defined here as dwarf stars cooler than 4500\,K.

 These plots indicate that M-dwarf stars have the highest values of MgI1.18 index while the highest values of SiI1.20 and SiI1.58 correspond mostly to giant stars. The exceptionally strong Si line strength observed in NGC\,1277 (SiI1.20 $\sim$ 1.8 and SiI1.58 $\sim$ 1.5) corresponds to a regime in the stellar library populated only by a few cool giant stars (grey and blue points), with no stars exhibiting higher Si absorption at lower temperatures (except a few M dwarfs and one carbon star in the case of SiI1.20). This suggests that the observed strength cannot be attributed to a particular stellar evolutionary phase (e.g. asymptotic giant branch stars in the case of CO band heads; \citealt{eftekhari2022}), as there is no population in the library with higher Si. Therefore, the Si enhancement in NGC\,1277 is best explained by a genuine increase in [Si/Fe], rather than being an artefact of a poorly modelled stellar phase.

We caution that the SPS models used are based on empirical stellar spectra of Milky Way stars, assuming similar abundance patterns and evolution in other galaxies such as NGC\,1277. Deviations from this—such as poorly modelled evolutionary phases or chemical peculiarities—could impact the interpretation of spectral features.

\section{Effects of IMF shape on line-strength indices}\label{app:secA2}

\begin{figure} 
        \centering
        \includegraphics[width=\columnwidth]{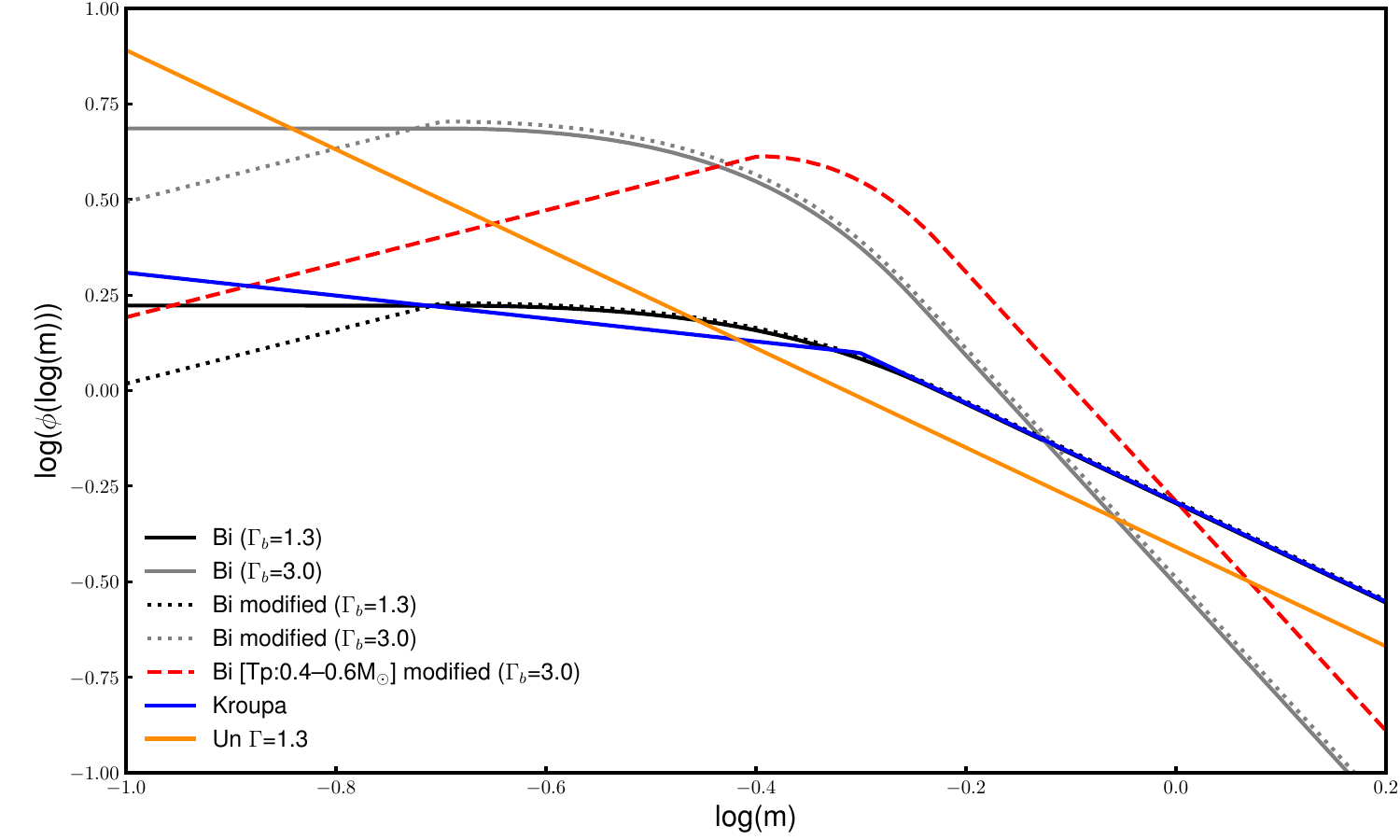} 
        \caption{Illustration of alternative IMF parameterisations used in the stellar population modelling. The standard Salpeter IMF is shown in orange. Solid black and grey lines represent bimodal IMFs with logarithmic slopes of 1.3 and 3.0, respectively. Dotted versions of these lines modify the low-mass slope below 0.2\,M$\odot$ to 0.3, mimicking the Kroupa IMF in that regime. The dashed red line corresponds to a bimodal IMF (slope = 3.0) with a reduced turning point mass range of 0.4–0.6~M$\odot$. Finally, the solid orange line also doubles as a unimodal IMF with slope = 1.3 for comparison. 
                }
        \label{fig:fig7} 
\end{figure}

To explain the higher line-strength indices observed in NGC\,1277 compared to massive ETGs, it is necessary to enhance the contribution of stars with the highest Si and Mg strengths while reducing the influence of stars with lower index values. This can be achieved by modifying the shape of the IMF. Figure~\ref{fig:fig7} illustrates various possible IMF shapes. For reference, we also show the Salpeter IMF and bimodal IMFs with slopes of 1.3 and 3.0. The bimodal IMF is parameterised as a two-segment function \citep{vazdekis1996}. Below 0.2M$\odot$ and above 0.6M$\odot$, the bimodal IMF is characterised by a power-law distribution. Between 0.2M$\odot$ and 0.6M$\odot$, it is a spline function that satisfies some boundary conditions. Below 0.2M$\odot$, the bimodal IMF follows one power-law slope of zero, and above 0.6M$\odot$, it follows a varying slope. 

As mentioned, M-dwarf stars have the highest values of MgI1.18 index. Therefore, increasing the slope of a bimodal IMF would enhance the number of M-dwarf stars with respect to giant stars, increasing the MgI1.18 absorption. In the case of Si indices, since most of the M-dwarf stars have lower values than giant stars, by increasing the slope of a bimodal IMF, the overall value of the Si indices decrease. These are shown quantitatively in Fig.~\ref{fig:fig6} where indices are measured on SSPs that are constructed based on the IMFs shown in Fig.~\ref{fig:fig5}. 

\begin{figure*} 
        \centering
        \includegraphics[width=0.9\textwidth]{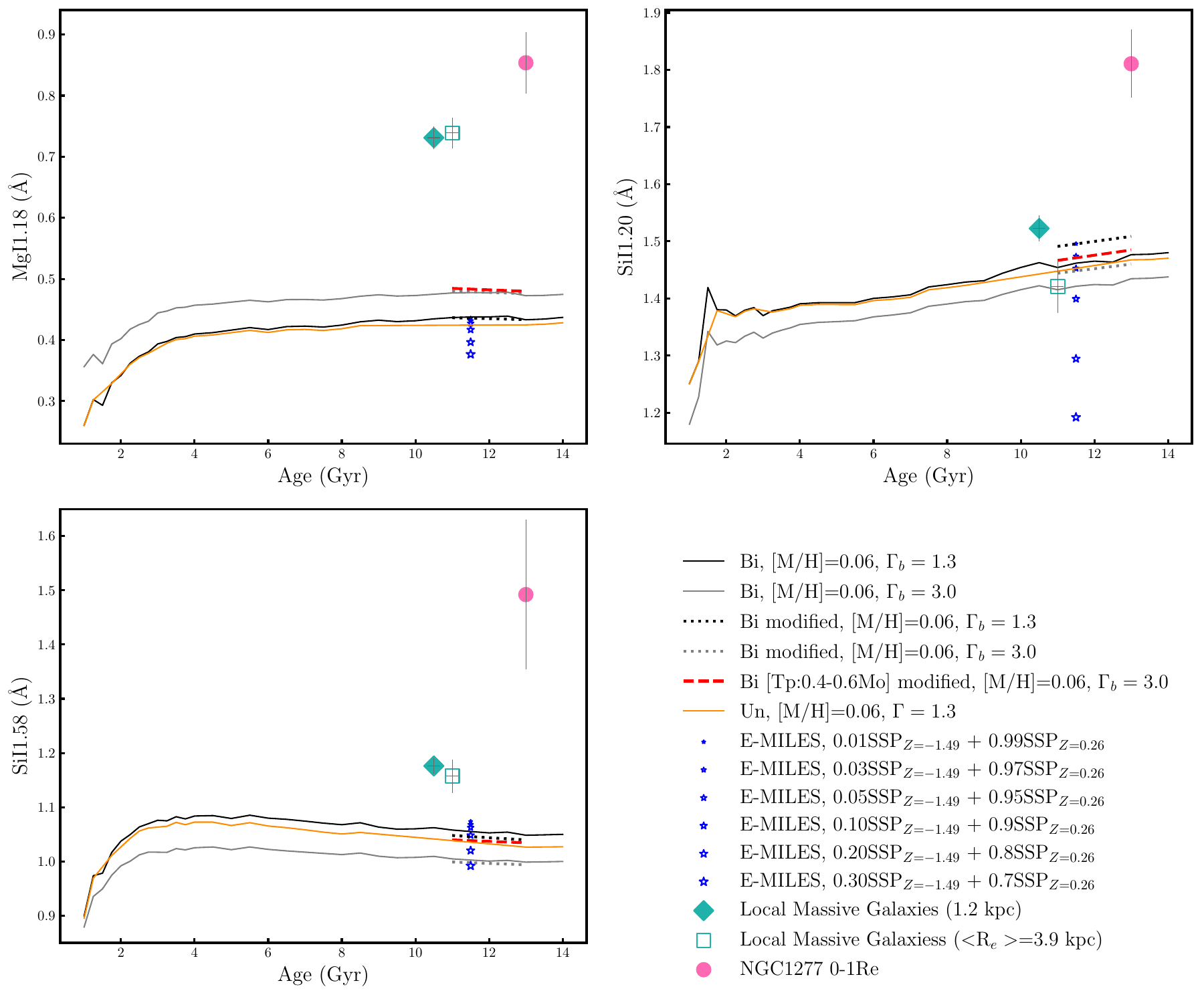} 
        \caption{Si and Mg line-strength indices as a function of stellar population age for different IMF shapes and stellar metallicity components. Each curve represents a SSP built using the IMF shapes from Fig.~\ref{fig:fig7}, with line styles and colours matched. Star symbols denote two-component SSP models that include a metal-rich population ([M/H] = $+$0.26~dex) combined with a varying fraction of a metal-poor component ([M/H] = $-$1.49~dex). Star sizes increase with the metal-poor fraction: 0\%, 5\%, 10\%, and 20\% by mass. We also show the impact of different aperture effects, with the two aperture extractions from massive ETGs shown as filled circles for central (1.2~kpc) apertures and open squares for spectra extracted within 1R$_e$. }
        \label{fig:fig8} 
\end{figure*}

We modified the definition of the bimodal IMF for low-mass stars by considering a slope of 0.3 for stellar masses below 0.2M$\odot$\footnote{This slope is chosen to be similar to the slope of the Kroupa IMF for very low-mass stars.}. This aimed to reduce the emphasis on the very cool stars that weaken the Si absorptions. The results are shown in Figs.~\ref{fig:fig7} and~\ref{fig:fig8}. The effect of this IMF shape on line-strength indices is almost negligible.

We also attempted to change the mass range of the turning point for the bimodal IMF. The turning point mass range was reduced from 0.6M$\odot$ to 0.4M$\odot$. The result for a bimodal IMF with a slope of 3.0 is shown in Figs.~\ref{fig:fig7} and \ref{fig:fig8}. In this case, only the contribution of M-dwarf stars that most affect the Si is emphasised, although the effect on the line-strength indices of Si is negligible. 

We also note that a unimodal IMF with the same slope as bimodal IMF with $\Gamma_{b}=1.3$ does not flatten at masses lower than 0.2M$\odot$, meaning that cooler dwarf stars (with lower MgI values) outnumber stars near the peak of the MgI strength. Consequently, the MgI1.18 line strength on SSPs built based on a unimodal IMF is similar to the ones built based on a bimodal IMF with $\Gamma_{b}=1.3$ (see the solid orange line in Fig.~\ref{fig:fig8}). All in all, none of these adjustments adequately address the discrepancies in Si indices. 

\section{Role of metal-poor populations}\label{app:secA3}

Further inspection of Fig.~\ref{fig:fig6} reveals a trend in dwarf stars (red symbols) where Si line strengths increase with higher [Fe/H]. This suggests that the elevated Si absorption in NGC\,1277 results from a lower fraction of metal-poor stars compared to the massive ETGs. Si, synthesised via the $\alpha$-process in massive stars, is released into the ISM by Type II SNe, enriching the gas that forms subsequent generations of stars ($\sim$200 Myr after the Big Bang and later).
    
Chemo-evolutionary population synthesis models \citep{vazdekis1996, vazdekis1997} suggest that massive ETGs contain two stellar populations: a dominant old, metal-rich population and a minor, metal-poor population formed during rapid early chemical enrichment. The enhanced Si indices in NGC\,1277 indicate a more rapid and efficient early star formation phase, leading to fewer metal-poor stars than in massive ETGs. We tested this hypothesis by constructing two-component SSP models, combining metal-rich ([M/H] = $+$0.26~dex) and metal-poor ([M/H] = $-$1.49~dex) populations. As shown in Fig. \ref{fig:fig7}, increasing the mass fraction of the metal-poor component (blue star symbols) affects the line strength indices. A 20\% metal-poor fraction lowers the SiI1.20 index to values seen in massive ETGs, but this level is inconsistent with optical and surface brightness fluctuation studies of massive ETGs, which limit this fraction to $\lesssim$5\%~\citep{vazdekis1997, blakeslee2001, ferreras2015, vazdekis2020}. Moreover, this adjustment does not reproduce the SiI1.58 index. 

\section{Robustness tests}\label{app:secA4}
\subsubsection{Index variations with aperture}

To study aperture effects, we repeated our analysis  using spectra extracted within one effective radius of the X-shooter individual galaxies (i.e. not in the stacked spectrum). This aperture plus the one that we used throughout the paper (matching the aperture for NGC 1277, i.e. 1.2 kpc) are shown in Fig.~\ref{fig:fig7}. We find that MgI1.18 and SiI1.58 values are indistinguishable within the measurement uncertainties. For SiI1.20, a lower value is observed in the $1R_e$ aperture, yet it remains significantly below the value measured for NGC\,1277. This confirms that the strong Si and Mg features observed in NGC\,1277 compared to massive ETGs are not the result of aperture effects.

\subsubsection{Abundance estimates with SPS models}

We repeated the abundance estimation using SPS models constructed based on the empirical X-shooter Spectral Library (XSL; \citealt{verro2022b}) to assess the impact of model systematics. XSL provides a wide wavelength coverage (3000–25,000~\AA) at a resolution of $R \sim 10,000$, enabling consistent modelling of both optical and NIR spectral features.

These models are available with two different sets of isochrones: Padova00 \citep{girardi2000} and PARSEC-COLIBRI (PC). The latter combines PARSEC v1.2S evolutionary tracks \citep{bressan2012, chen2014, chen2015} with COLIBRI thermally pulsing asymptotic giant branch models \citep{marigo2013, rosenfield2016, pastorelli2019}, improving the treatment of the thermally pulsing asymptotic giant branch phase, crucial for accurate modelling of the NIR. They span a wide range of ages from 1 to 13.5~Gyr and metallicities from [M/H] $=–$1.5 to $+$0.3, adopting a Kroupa-like IMF.

In this study, we exclusively used the XSL SSP models constructed with the PC isochrones. Our selected model grid spans five ages (1.0, 2.5, 5.0, 10.0, and 12.6~Gyr) and three metallicities ([Fe/H] = 0.0, $+$0.1, $+$0.2). 

The overall trends in [Si/Fe] and [Si/Mg] remain stable. Although the XSL models produce slightly different absolute index strengths due to differences in spectral library and IMF, the relative differences between NGC\,1277 and massive ETGs persist: the [Si/Mg] ratio remains significantly elevated in NGC\,1277 (centred at $\sim$0.67\,dex) compared to massive ETGs (centred at $\sim$0.16). This further confirms that the Si-enhancement in NGC\,1277 is not an artefact of the adopted stellar population model.

\subsubsection{Optical estimates of Si and Mg}

To assess the consistency of our NIR-based abundance results, we analysed optical spectra of NGC\,1277 and the stacked sample of massive ETGs. The optical spectrum of NGC\,1277 was obtained with the ISIS spectrograph on the William Herschel Telescope (WHT) (see \citealt{trujillo2014} for observational details). Spectral fitting was carried out using both full index fitting and full spectral fitting methods, following the procedures outlined in \cite{labarbera2025}.

Each fit incorporated a two-component stellar population model (2SSP), consisting of an old and a younger SSP, to account for potential age variations. Abundance ratios of several elements (Mg, Na, C, Ca, N, Ti, Si, and O) were simultaneously fitted using response functions from \cite{conroy2018} theoretical stellar population models, with the same methodology across NGC\,1277 and stacked spectra of massive ETGs.

This analysis yielded four independent [X/Fe] estimates per element (2 fitting methods × 2 isochrones), with means and standard deviations computed to quantify uncertainties. For Si in particular, the [Si/Fe] in NGC\,1277 was found to be modestly enhanced (by $\sim$0.05~dex) relative to massive ETGs, consistent in sign but smaller in amplitude than our NIR-derived estimate.

As an additional optical consistency check, we compared two full-spectrum fits: one in which [Si/Fe] was allowed to vary freely, and one in which [Si/Fe] was fixed to a high value of +0.6~dex, representative of the NIR-inferred enhancement, while the remaining stellar-population and abundance parameters were refitted. The free-[Si/Fe] fit favours a lower abundance, around [Si/Fe] ($\simeq$~0.2) dex. The fixed-high-[Si/Fe] model provides a somewhat poorer fit and introduces localised residuals in some spectral regions, particularly around blended CN and metal-line complexes. However, the mismatch is not severe, with residuals remaining comparable to the observational uncertainties. We therefore interpret the optical spectrum as favouring a lower absolute [Si/Fe] value, while not decisively ruling out the higher value inferred from the NIR.

As a complementary theoretical sensitivity check, we also inspected theoretical optical SSP spectra, which we computed with the ISPy3 code \citep{larsen2020} using two different Si abundances, [Si/Fe] = +0.3 and +1.0. Both SSPs assume an old age of 12 Gyr ($\alpha$-enhanced BaSTI  isochrones; \citealt{pietrinferni2004}), a bottom-heavy IMF, and a metal-rich baseline composition of [M/H] = +0.2. The abundances of O, Ne, Mg, S, Ar, Ti, and Na were fixed at +0.3 dex, while Ca was kept solar. The models were computed using MARCS atmospheres for the coolest stars, with (T$_{\rm eff}$<4000,{\rm K}), and ATLAS12 atmospheres for the remaining stars, allowing the atmospheric structure and opacities to be computed consistently for the adopted abundance mixture. After smoothing the SSPs to the velocity dispersion of NGC\,1277, ($\sigma$ = 440,{\rm km,s$^{-1}$}), we find that the strongest optical response to increasing Si occurs in the blue, particularly around SiI4102. For this feature, the index increases from 1.07~\AA\ for [Si/Fe] = +0.3 to 1.79~\AA\ for [Si/Fe] = +1.0. However, the observed difference between NGC\,1277 and the X-shooter galaxy (XSG) stack is very small, (0.831$\pm$0.029)~\AA\ versus (0.807$\pm$0.022)~\AA. Conversely, SiI4513 is slightly stronger in NGC\,1277 than in the XSG stack, (2.078$\pm$0.027)~\AA\ versus (1.990$\pm$0.027)~\AA, but the SSPs predict only a weak response in the opposite direction, decreasing from 2.47~\AA\ to 2.40~\AA. This confirms that the optical Si-sensitive features do respond to changes in Si abundance, but not in a clean or uniquely interpretable way at the velocity dispersion of NGC\,1277.

The discrepancy likely stems from a critical limitation of optical spectra at the velocity dispersions typical of massive galaxies: the absence of clean, isolated features with strong sensitivity to Si. While Si-sensitive indices are formally included in optical index libraries (e.g. \citealt{serven2005}), they are dominated by blended contributions from multiple elements, particularly in smoothed spectra with high velocity dispersion. Specifically, the commonly used SiI3905~\AA\ line is significantly blended with CH molecular features, especially problematic in cool giants \citep{cayrel2004}, while the SiI4102~\AA\ line lies on the wing of the strong H$_{\delta}$ absorption, further complicating accurate abundance determinations \citep{cohen2004}. We also inspected the red Si-sensitive windows at 5661–5671, 5685–5695, 5767–5777, 6150–6160, and 6232–6250 Å. However, in the NGC1277 spectrum, several of these regions are affected by broad continuum distortions and residual structures that are not present in either the XSG stack or the stellar population models. We therefore did not use these narrow optical windows to derive additional quantitative Si abundance constraints.

These limitations likely contribute to contradictory results in the literature. For example, \cite{worthey2014} report [Si/Fe] abundances that are systematically higher than [Mg/Fe] in their galaxy sample. However, their estimates rely heavily on blue SiH molecular features, which are not well isolated in index-based analyses and are difficult to model reliably with synthetic spectra. As a result, their [Si/Fe] values are particularly sensitive to subtle systematics in the input data, leading to potentially unstable or uncertain abundance determinations. In contrast, \cite{conroy2014}, employing full spectral fitting with flexible IMFs and high-resolution synthetic models, often report higher [Mg/Fe] than [Si/Fe] (e.g. their Fig. 19). Yet because full spectral fits optimise the entire spectrum simultaneously without directly evaluating the quality of Si-dominated regions, it is not always clear how well constrained the [Si/Fe] values are in these analyses.

Altogether, these uncertainties emphasise the limitations of optical diagnostics for silicon. The optical analysis recovers a relative Si enhancement in NGC\,1277 compared to massive ETGs, but with a smaller absolute amplitude than inferred from the NIR. We therefore regard the optical–NIR difference as a systematic uncertainty in the absolute abundance scale. In contrast, the NIR features used in our analysis are stronger, less blended, and more sensitive to Si abundance variations. These advantages make the NIR regime the preferred spectral window for tracing Si enrichment in stellar populations.

\section{Elemental abundance trends: [Si/Fe] and [Mg/Fe] versus progenitor mass}\label{app:sm_f}

\begin{figure}
        \centering
        \includegraphics[width=\columnwidth]{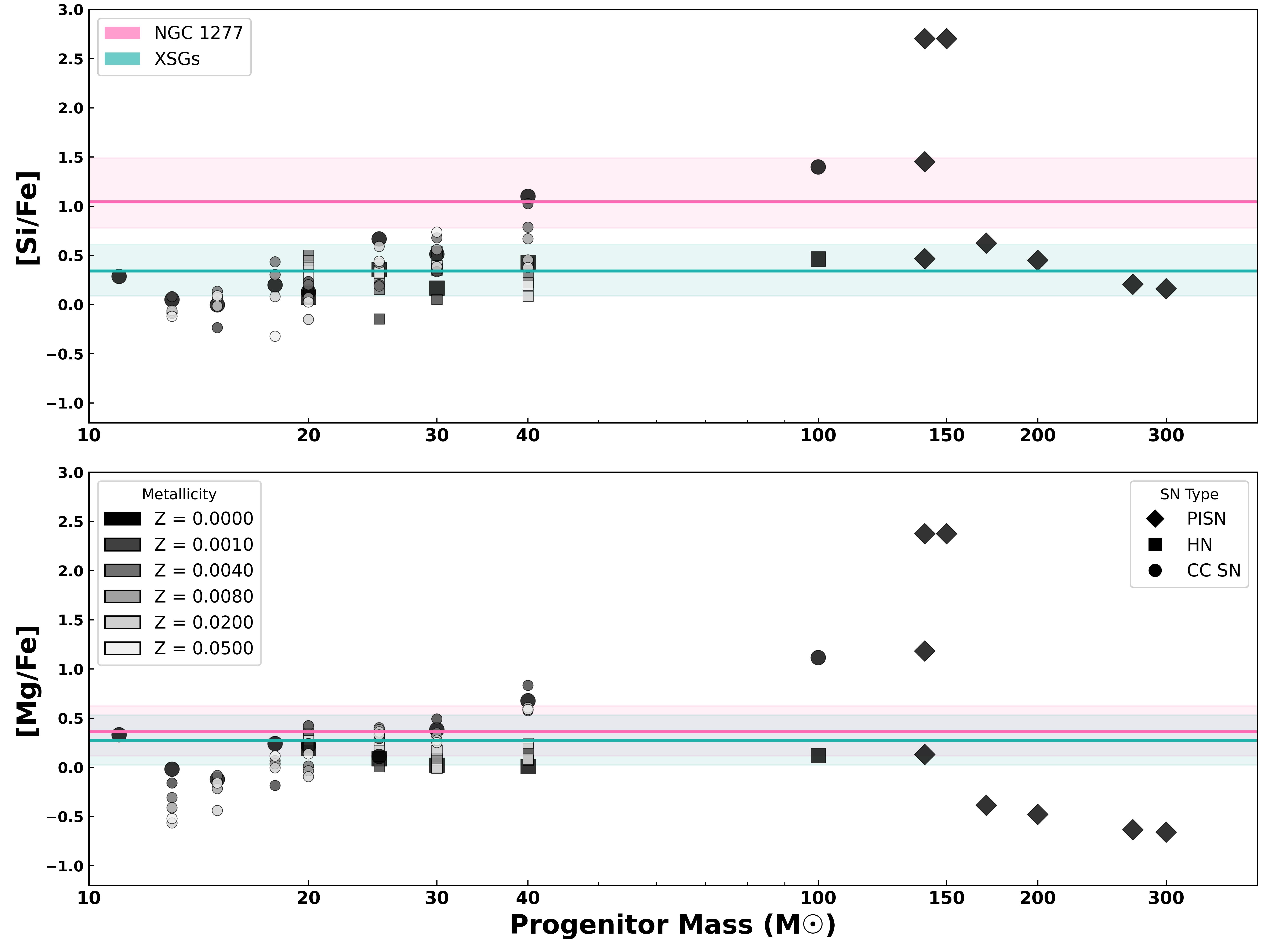}
        \caption{Measured galaxy abundances compared with predicted [Si/Fe] and [Mg/Fe] yields from SN models as a function of progenitor mass and metallicity. As in Fig.\,\ref{fig:fig2}, circles indicate CCSNe, squares HNe, and diamonds PISNe. Symbol colours indicate metallicity from Z=0 (black) to Z=0.05 (light grey). The shaded bands show the 68\% credible intervals of the inferred abundances for NGC 1277 (pink) and the massive ETG sample (cyan), with solid lines marking the corresponding posterior medians. While the [Mg/Fe] abundance is broadly consistent with several enrichment channels, the high [Si/Fe] inferred for NGC\,1277 is best matched, within the adopted yield grid, by very massive, metal-poor progenitors, including PISN models.
                }
        \label{fig:fig5} 
\end{figure}

Complementing Fig.~\ref{fig:fig2} in the main text, we present in Fig.~\ref{fig:fig5} the predicted [Si/Fe] and [Mg/Fe] values as a function of SN progenitor mass and metallicity. These plots provide the underlying elemental abundance trends from which the [Si/Mg] ratio is derived, enabling a more complete interpretation of our results.

One can see that the [Mg/Fe] values for both galaxy samples fall within the range of predictions from all types of SNe . The [Mg/Fe] values are also consistent with estimates from optical studies \citep{trujillo2014}, with NGC\,1277 showing a slightly higher [Mg/Fe] enhancement than massive ETGs. In contrast, the [Si/Fe] panel shows a clear divergence: the high [Si/Fe] values inferred for NGC\,1277  are best matched by models of metal-free SN progenitors, in particular a subset of PISNe with initial masses above $\sim$150\,M$_\odot$. A few of these PISN models predict [Si/Fe] values even higher than those observed in NGC\,1277.

These results reinforce the interpretation that the unusually high [Si/Mg] ratio in NGC\,1277 originates from the combined effect of elevated [Si/Fe] and modest [Mg/Fe], both shaped by enrichment from early, massive, and likely metal-free progenitors.

\end{appendix}
\end{document}